# Real-time microscopic view of the relaxation of a glass

Marta Ruiz-Ruiz[1,2], Ana Vila-Costa[1,2], Tapas Bar[2], Cristian Rodríguez-Tinoco[1,2,*], Marta Gonzalez-Silveira[1,2], Jose Antonio Plaza[3], Jorge Alcalá[4], Jordi Fraxedas[2], Javier Rodriguez-Viejo[1,2,*]

**Affiliations:**

[1]*Departamento de Física. Facultad de Ciencias, Universitat Autònoma de Barcelona, 08193, Bellaterra, Spain*

[2]*Catalan Institute of Nanoscience and Nanotechnology (ICN2), CSIC and BIST, Campus UAB, Bellaterra, 08193, Barcelona, Spain*

[3]Instituto de Microelectrónica de Barcelona, IMB-CNM (CSIC), Esfera UAB, Campus UAB, Cerdanyola, Barcelona, Spain

[4]InSup, ETSEIB, Universitat Politècnica de Catalunya, 08028 Barcelona, Spain

**Abstract**

The understanding of glassy dynamics above the devitrification temperature of a glass remains poorly understood. Here, we use real-time AFM imaging to build a spatio-temporal map of the relaxation dynamics of a highly stable glass into its supercooled liquid. This new methodology enables a direct visualization of the progression of the liquid phase and clarifies and quantifies the presence of localized fast mobility regions separated by giant length scales. Our data permit to establish a clear correlation between dynamic length and time scales in glasses. This approach may also be applicable to unveil the microscopic structure and dynamics of other glass forming systems with much shorter length and time scales, including liquid-cooled glasses.

Corresponding authors: javier.rodriguez@uab.cat; cristian.rodriguez@uab.cat

Glasses are non-equilibrium materials arrested during cooling in metastable configurations. The temperature that marks the transition from the ergodic state, the liquid, to the non-ergodic one, the glass, is known as the glass transition temperature, $T_g$. A hallmark in liquid and glassy dynamics is the recognition that dynamic heterogeneities (DH) are at the core of the slowing down of the dynamics and are responsible for the glass transition and its temperature dependence[1,2]. DH are characterized by clusters of regions of atoms/molecules with correlated mobility that grow as the temperature is decreased. The existence of these clusters of mobility has been inferred from previous experiments[1,3,4], but it is experimentally challenging to identify them or their length scales, although it is recognized that the broad distribution of time scales in a glass can be associated with the presence of mobile and immobile regions[5,6]. It is generally accepted that the relaxation of the glass into the supercooled liquid (SCL) at the devitrification temperature on heating, $T_{on}$, happens through a gradual softening of the glass across its entire volume with a correlation length of few nm. However, we have recently shown this view is not unique and can be changed by accessing a temperature/time regime above $T_{on}$ where glasses transform into the liquid by the formation of localized regions of liquid within a glassy matrix[7], even though the associated length scale has not been directly measured by experiments yet[8–10]. Vapor-deposited stable glasses appear as model systems to explore this scenario[11–14] and recent simulations point in this direction[15].

Since direct spatial visualization of equilibrated regions above $T_{on}$ is extremely challenging due to the subtle structural changes over very small distances, we follow a different strategy based on the local mechanical instabilities that the liquid regions generate on a rigid ultrathin layer grown on top of the glass. Well defined surface undulations can be produced in several ways[16,17], leading to self-organized wrinkling patterns with a rich variety of morphologies depending on the particularities of the system[18,19]. One approach to induce surface wrinkling is by capping a soft thin-film material (typically a polymer, but also a small-molecule organic glass) spin-coated on a rigid substrate by a thinner film of a metal layer[17] or by another organic layer with a higher $T_g$[20]. The instability can be initiated by annealing the organic material into the rubbery state where wrinkling across the whole surface appears due to the development of compressive stresses induced by differences in thermal expansion coefficients[21,22]. Interestingly, wrinkling can be locally induced at specific sites of the film by localized surface modifications using focused ion beams[23], swelling-induced stress through toluene absorption[24], or local heating with an external source, such as a laser[25].

Here, we take advantage of a previously established protocol to induce bulk 'melting' in thin film glasses[7,26,27] by capping the organic film (N,N'-Bis(3-methylphenyl)-N,N'-diphenylbenzidinem,

TPD, $T_{g1}$ = 333 K measured at 10 K/min) with two ultrathin layers of an organic glass with a higher $T_g$ (Tris(4-carbazoyl-9-ylphenyl)amine, TCTA,$T_{g2}$ = 428 K, 10 K/min). The mechanical instabilities produced during isothermal treatments above $T_{g1}$ are measured in-situ using Atomic Force Microscopy (AFM) and optical microscopy (OM). We address the continuum mechanics finite element (FE) modelling of the wrinkling phenomenon to demonstrate that the local surface wrinkles appearing at $T_{ann} > T_{g1}$ (TPD) are induced by the equilibration of the intermediate TPD layer at localized spots. The observed patterns are compatible with a time-dependent number of initiation sites that appear to propagate radially. With this data we construct a spatio-temporal, microscopic map of the relaxation of a molecular glass, delivering a clear picture of the heterogeneous devitrification dynamics of glasses with enhanced stability. Imaging by AFM or OM provides a direct visualization of giant lengths scales between fast mobility regions and the growth of the liquid by dynamic facilitation. This methodology also offers an original approach to study the glass transition from a localized and novel perspective.

A sketch of the trilayer organic structure grown on a 500 μm thick Si substrate with its native oxide (surface RMS = 0.13 nm in 1x1 μm$^2$) together with the thermal protocol is shown in Fig. 1a,b. The TPD layer is vapor-deposited at a substrate temperature of $0.85T_{g1}$ (285 K) and has a fictive temperature[28], $T_f$=292 ± 2 K, 39 K below its $T_{g1}$. Details of the growth of the organic layers and the experiments are given in Methods. After growth, the thin film stack (13 nm TCTA/63 nm TPD/13 nm TCTA/Silicon, Fig. 1a) is taken to an AFM or optical microscope equipped with a heating stage. Most results involve AFM due to its superior spatial resolution. In the AFM the temperature is raised to a working value of 349 K, $T_{g1}$+16 K, in sequential steps as shown in Figure 1b to allow for equilibration of the AFM setup (Methods). At 349±1 K the transformation time of TPD into the SCL is around 270 min ($t_{trans} \approx 2x10^6 \tau_\alpha$, being $\tau_\alpha$ the relaxation time of the equilibrated liquid at that temperature)[29], taking t=0 s as the time at which this temperature is reached. This time interval offers a convenient scale to follow the kinetics of the transformation in real time, given that each AFM scan takes around 4 min to complete (see Methods). Figure 1c shows some representative in-situ AFM images obtained at T= $T_{g1}$+16 K in the same spatial location for different times. The surface of the trilayer was initially very smooth (RMS roughness below 0.3 nm in 1x1 μm$^2$) and did not show any surface undulation before reaching the annealing temperature. Figure 1b bottom represents the apparent height of the surface topography across the line scans marked in the AFM images (Methods). The images reveal the emergence of surface corrugation that initially appears in the form of Gaussian-type protuberances with aspect ratios around 100, that is few nm in height and few hundred nm in width and develop over time to localized wrinkle patterns that grow radially from the initial seed. A set of AFM (OM) images of the transformation of a 13/63/13 nm trilayer are shown as movies in Supplementary Videos S1 and S2. We note that the onset of the perturbation could not be identified by AFM due to the time scale of the measurements, nor by OM due to its reduced spatial resolution, so these measurements are not sensitive to the initial formation of the liquid droplet within the film and only monitor its appearance when the liquid region is larger than the thickness of the middle layer, see schematics of Fig. 2a. The peak-to-peak amplitude of the central part evolves from 2-3 nm of the initial seeds to around ~30 nm (peak-to-valley) at the end (Fig. 1c), although the precise dimensions depend slightly on the selected wrinkled pattern. The dominant wavelength of the fully transformed sample, obtained from the power spectrum analysis of the FFT image, Fig. 1d, amounts to 912 nm (Methods). The AFM images in Fig. 1 and supplementary videos 1 and 2 show two main features: the appearance of new local corrugations of the trilayer as time evolves and their radial growth. Corrugation is irreversible and maintains its structure over months at room temperature. The observed patterns are not associated with crystallization, XRD as well as TEM imaging of the fully corrugated sample after the thermal treatment show that the trilayer maintains its amorphous structure (Supplementary Fig. 1). Capping the TPD middle layer at both sides is required to avoid surface induced transformation and to observe localized patterns evolving with time. TPD samples grown directly on weakly interacting native oxide/Si substrates and capped only on top by TCTA do show wrinkling at the whole surface from the beginning of the experiment due to the transformation of TPD into the SCL by a propagation front that starts at the TPD/SiO$_2$ interface (Supplementary Fig. 2). This behaviour is consistent with the thermal wrinkling commonly observed in metal/polymer bilayers[30].

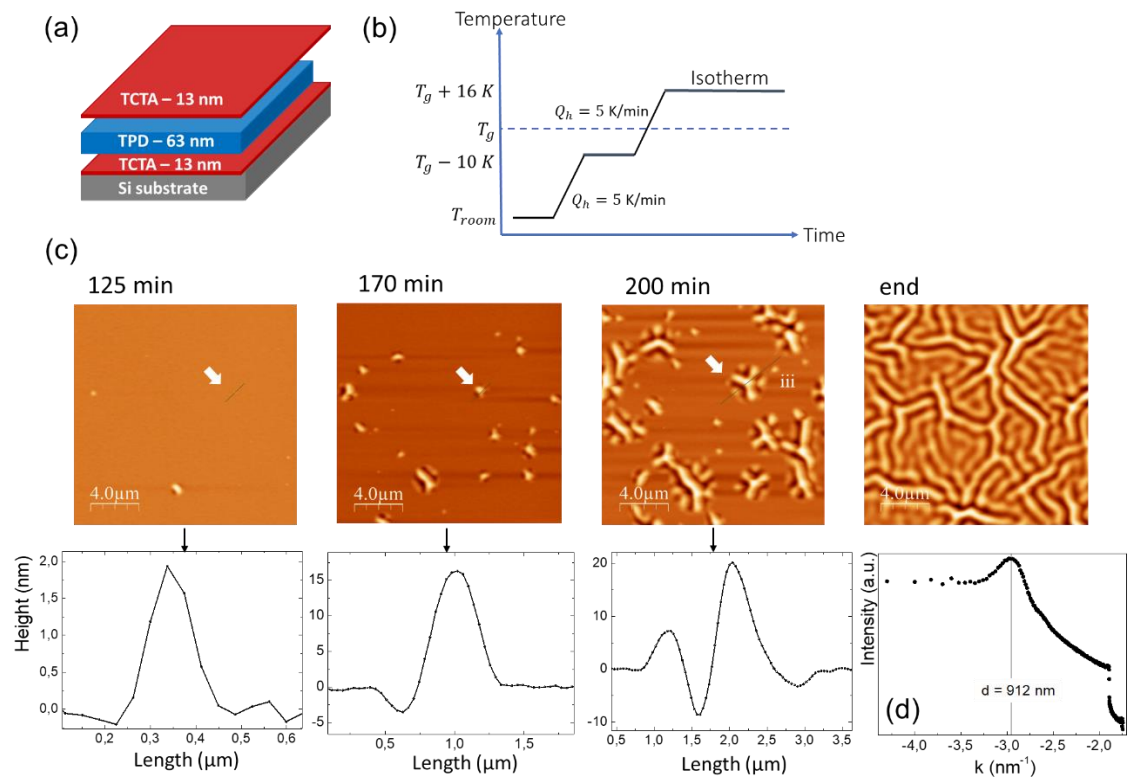

**Fig. 1. Representative images of the highly heterogeneous transformation of a stable TPD glass.** (a) Sketch of the trilayer structure and (b) of the annealing protocol in the AFM setup. (c) AFM snapshots obtained at 349 K after time t showing the formation and progression of local wrinkles induced by the transformation of the middle TPD layer into a SCL. Bottom graphs show the line profiles of the selected regions. A white arrow marks the pattern considered. (d) Power Spectral Density distribution of the fully transformed sample indicating the prominent wavelength of the wrinkled distribution.

## Mechanical analysis of wrinkle formation

Continuum mechanics FE simulations are used to correlate the formation of surface undulations with the existence of supercooled liquid regions within the intermediate TDP layer. A distinctive feature of the investigated mechanical instabilities compared with most previous thermal wrinkling studies is their local character, where the softened (liquid) sites slowly populate the interlayer. While the three-layer thin film system thus behaves in accordance with linear thermoelasticity, cylindrical SCL regions arise in the middle TPD layer after some time above $T_{g1}$. Surface undulation patterns are then induced due to the transfer of the compressive thermal loads from the TPD interlayer to the stiffer and much thinner TCTA top layer, which remains in its glassy state. This is a mechanically unstable phenomenon triggered under the ~33-fold higher elastic modulus and ~88-fold smaller thermal expansion coefficient of the Si substrate compared to the organic layers. These surface undulation processes are schematically illustrated in Fig. 2a, resulting in the experimentally observed wrinkled patterns. Concentrically ridged undulations are also produced as a function of the relative layer thicknesses, see Supplementary Fig. 3.

The onset of the first surface undulation (single ridge configuration) is concomitant to the vertical expansion of the cylindrical SCL regions due to density ($\rho$) differences between liquid and glass ($\rho$(UG) 2.8% higher than $\rho$(SCL) at $T_{g1}$ +16 K[31,32] and the compression induced by the surrounding TPD glassy interlayer. The capacity of the cylindrical SCL regions to withstand shear stresses at the outermost regions close to the top TCTA layer is then regarded as an essential element to the attainment of the multiple periodic undulations associated with the mechanical instability of the thin film system. The sudden inception of fully wrinkled surfaces with the partly capped, two-layer system (Supplementary Figure 2) is indicative in that this capacity is maintained across larger surface areas. Arguably, some viscous relaxation of the shear stresses is finally envisioned to occur at $T_{g1}$ +16 K, tentatively reducing the heights of the undulations.

In our FE simulations, we consider that the softened TPD interlayer is a dense liquid that exhibits solid-like properties below a certain elastic limit. Since we are essentially interested in assessing the onset of the surface undulations under sustained shear, the cylindrical SCL regions are envisioned to deform through linear elasticity with discretionary levels of compressibility and small elastic modulus. The softened interlayer is alternatively assumed to deform under the non-linear Neo-Hookean hyperelastic model, so as to capture a possible decrease in the initial elastic modulus at small strain levels (Methods). Comparisons between the experimental and computational results are shown in Figs. 2b-d. A cross-section scanning electron micrograph showing the surface undulation of a trilayer is shown in Supplementary Fig. 8b.

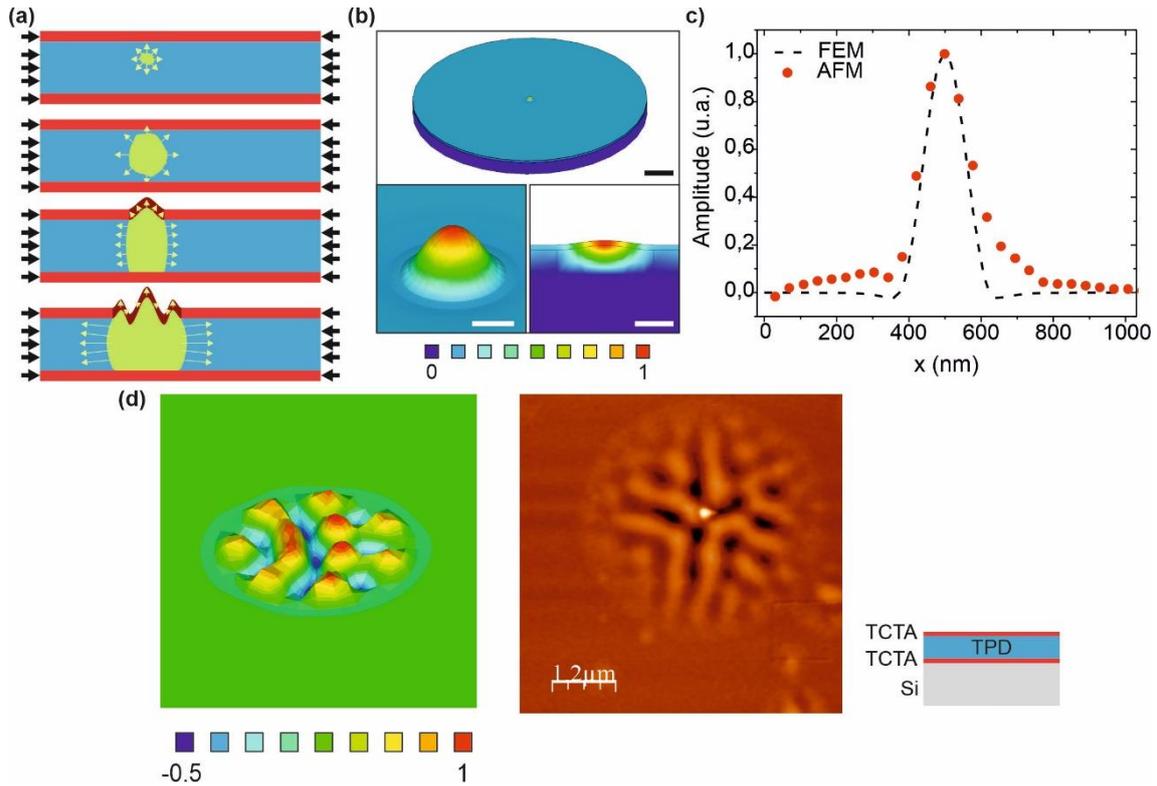

**Fig.2 Surface undulation patterns and their modelling *via* FE simulations.** (a) Schematic showing the growth of an SCL region (in green) from a nucleation site within the TPD interlayer towards the top and bottom TCTA layers (growth direction marked with green arrows). This results in the development of the first surface undulation triggered under the applied thermal stresses (marked with black arrows). Notice the periodic onset of further secondary undulations as the cylindrical-shaped SCL region extends radially and mechanical instabilities develop. (b) Simulated normalized out-of-plane displacement (top) isotropic view of the whole simulated structure, (bottom, left) isotropic view a zoom of the first surface undulation region of diameter $\theta \sim 250$ nm within the TPD interlayer and (bottom, right) a cross section of the (bottom, left). The neo-Hookean model with material parameters $C = 3.71 * 10^6$ Pa and $d = 5.58 * 10^{-8}$ Pa$^{-1}$ is assumed in the simulations. (c) Comparison between the simulation results and the experimental AFM measurements concerning the shape of the first undulation during the early propagation stages of the liquified front. (d) Comparison between a simulated wrinkled pattern of $\theta \sim 1000$ nm under the assumption of the neo-Hookean model with material parameters $C = 3.71*10^6$ Pa and $d = 5.58*10^{-8}$ Pa$^{-1}$, and the AFM image of a typical pattern in 13/63/13 nm trilayers.

**Equilibration of liquid regions: Giant length scales.**

The images of Figure 1c and Supplementary Videos 1 and 2 offer a detailed view of the evolution of the glass upon a temperature jump above $T_{g1}$, in which localized regions within the TPD glass having higher mobility evolve towards the liquid state —a process referred to as softening and equilibration, subsequently consuming the glassy environment by dynamic facilitation. These images clarify previous experiments and theories about the bulk transformation of stable glasses[11–13,33]. These studies, through indirect measurements, suggested that bulk melting on these glasses proceeds through a 'nucleation and growth'-type process with giant length scales between the initial patches of liquid. Here, we assume the first regions equilibrating into the liquid correspond to those with the highest mobility and therefore we interpret the emergence of newly equilibrated regions at different times as clear evidence of dynamic heterogeneity. Our results confirm that DH in ultrastable glasses is characterized by giant length scales of many microns between fast mobility regions. Besides these localized spots, we also have indirect evidence that the entire glass matrix slowly softens during the annealing above $T_{g1}$, but to the most part without reaching equilibrium. Calorimetric measurements show the change in the position of the endothermic overshoot of the highly stable glass matrix during the annealing treatment (ref. [10] and Supplementary Fig. 4).

In Fig. 3 we represent the time evolution of the heterogeneous, localized softening obtained from AFM. Figure 3a plots the cumulative number of regions that enter the regime of equilibration and become liquid-like per unit area (black symbols). The inset is a plot of the newly developed liquid regions per unit area (black circles), and normalized by the area of the untransformed fraction (blue symbols) as a function of time, (see Supplementary Information). This representation highlights an acceleration of the number of local regions entering the liquid regime over time. We foresee two possible scenarios to explain this behaviour: i) acceleration originates from the softening of the matrix, or ii) it is directly related to the intrinsic shape of the distribution of relaxation times of the UG at this temperature, or to some mixed effect. In Fig. 3b we plot a 2D spatial map of the sample with the position of the equilibration spots as they appear. The colour scale marks the time from the appearance of the first liquid regions.

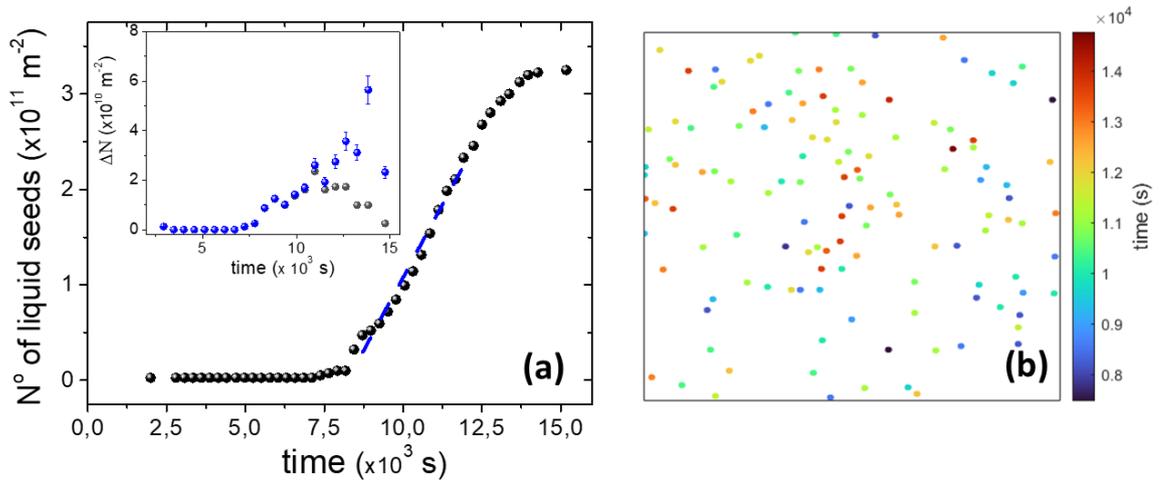

**Fig. 3. Formation of liquid seeds in the TPD interlayer during annealing above T_g.** (a) Evolution of the cumulative number of nuclei per unit area. (inset) number of new nuclei that appear in each AFM image per unit area (black symbols) and per unit area of untransformed glass (blue symbols). The error bars in the inset reflect the uncertainty in the transformed fraction; (b) spatial map at 90 % of transformation obtained from AFM images showing mobility regions where glass relaxes directly into the liquid. Colour code reflects the time of formation; t=0 marks the time at which the temperature reaches 349 K. The first soft spots appear after 2 h.

An initial period of about 2h without any noticeable surface feature can be clearly discerned in Fig. 3a. This long transient is related to the high stability of the glass, that is some regions attain liquid mobility within this long time. To some extent the large induction period could also be associated to a slower growth velocity of the minuscule initial seeds localized inside the TPD matrix before they reach the TCTA capping layers due to pressure effects originating from density differences between the UG and the SLC above $T_{g1}$, as reported recently in a numerical simulation study[15]. This effect may have a minor influence here because of the small layer thickness but could become important for thicker films. In the time interval between 0.9-1.25x10$^4$ s, the evolution of the number of liquid seeds (Fig. 3a) can be roughly approximated by a linear function (dashed blue line in Fig. 3a) giving an average 'formation frequency' ν≈6±2x10$^7$ nuclei/m$^2$s. This magnitude translates into a mean separation distance of liquid droplets appearing in 1 s per unit area of ≈130 μm, using $\xi(T) = \left( \nu \cdot (1s) \right)^{-1/d}$, being $d$ the dimensionality of the system. Experimentally, the giant length scale of the bulk transformation was previously indirectly inferred from the lower bound of the crossover length between front propagation and bulk melting, as well as from calorimetric measurements that identified the appearance of isolated liquid regions[8,9]. Length scales of the order of 1-4 μm have been suggested. To compare with previous estimates of the cross-over length in thin films, it is necessary to account for the time dependence. That is, in 3000 s the average distance between localized liquid regions roughly equals 2 µm (under the assumed $v=6\times10^7$ nuclei/m$^2$s (Fig. 3a)), which falls within the same order of magnitude as the ≈1.2 µm distance travelled by the front at $T_{g1}$ + 16 K in the same time scale (considering $v_{front}$≈0.4 nm/s, as discussed below). This confirms the length scale from the previous crossover experiments[8,9]

**Relaxation map of the glass**

We also follow the time evolution of the size of the liquid regions both from AFM and OM images (Fig. 4 and Supplementary Fig. 5). A strong indicator that the wrinkles originate from the presence of liquid in the TPD layer is found in the radial propagation velocity of propagation of the instability region. We have analysed many individual spots, such as those shown in Fig. 4a, rendering an average radial propagation speed at $T_{g1}$ + (16 ± 1) K of 0.3 ± 0.1 nm/s. This is consistent with the growth front velocity of the SCL TPD at this temperature, <v> = 0.4 ±0.1 nm/s, evaluated independently through extrapolation of high and low-temperature calorimetric data (see Suppl. Fig. 6). Thus, the wrinkle patterns propagate laterally at the speed of the liquid front, confirming previous hypothesis[34]. Although the liquid signature on the mechanical instability looks radial on average, we cannot infer if mobility locally spreads isotropically or through string-like, fractal, dynamics. From the wrinkled area we can estimate the liquid transformed fraction that follows a sharp sigmoidal shape, as shown in Fig. 4b, most probably due to the continuous formation of new liquid seeds and its acceleration as time evolves. Figures 4c and 4d represent spatio-temporal maps of the evolution of the liquid including its progression by dynamic facilitation taken at transformed liquid fractions X=34% and 90%, respectively, and can be interpreted as relaxation maps of the glass. As time progresses, regions of fast mobility spatially separated by giant length scales equilibrate towards the liquid and they ignite adjacent regions that also become more mobile by a facilitation mechanism. A direct comparison between the maps at 90% of the transformation of Figs. 3b and 4d evidence that most of the glass transforms into liquid by dynamic facilitation, suggesting that a large fraction of the as-prepared glass consists of highly stable, inactive regions which may only transform from adjacent zones that have become liquid. Remarkably ≈10-15% of the glass has still not transformed after $1.5\times10^4$ s at $T_{g1}$ + 16K, (darker brown areas in Fig. 4d), indicating that these regions exhibit relaxation times above ≈$2\times10^6$ $\tau_\alpha$. The mechanism described above is consistent across a range of annealing temperatures above $T_{g1}$ and thickness of the TPD layers (60-400 nm), as briefly shown in Supplementary Fig. 7, 8 and 10, but we leave a complete study for future work. We also note the dynamics can be slightly influenced by surface defects (or increased roughness of the substrate) as shown in Figs. S9 and S10.

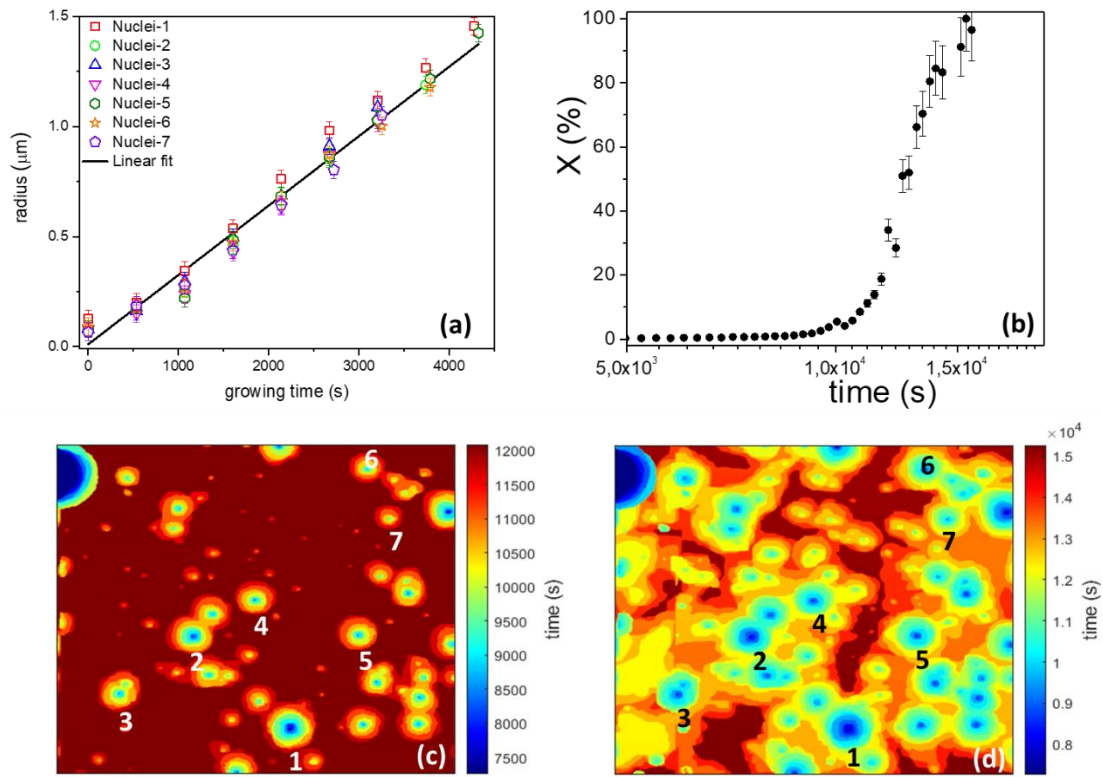

**Fig. 4. Relaxation maps of the stable TPD glass showing the impact of dynamic facilitation.** (a) Growth of several liquid regions as a function of time at $T_{g1}$ + 16 K. Growing time is normalized to the initial time for growth of each nucleus. The error bars reflect the standard deviation for each measurement; (b) Transformed fraction of liquid versus time. The error bars show the uncertainty in evaluating the transformed fraction. Spatio-temporal maps of the relaxation of the glass including lateral progression of the liquid by dynamic facilitation at (c) 34 % and (d) 90 % of transformed fraction. Colour scale marks the time evolution. The nuclei evaluated in figure 4a are indicated by numbers.

**Conclusions and outlook**

We have reported a clear, direct and real-time visualization of the spatio-temporal dynamics of the devitrification of an ultrastable glass by AFM and optical measurements. Our data unambiguously demonstrates the correlation between time and length scales in glasses. Glasses with ultraslow average dynamics and huge transformation times, $\tau_{glass} \approx 10^6 \, \tau_\alpha$, exhibit giant length scales, measured as the average distance between patches of fast mobility that equilibrate into the liquid, of the order of hundred microns at $T_{g1}$ + 16 K. This giant length scale paves the way to manipulate the properties of ultrastable glasses by nanostructuration, opening an interesting avenue to further influence the dynamics. We also anticipate that reducing the stability of the glass (i.e., its time scale) will also result in a reduction of the length scale. By using the mechanical instabilities described here it should be possible to test the existence of heterogeneity on a smaller time and spatial scale by using ultrathin capped layers of liquid-cooled glasses, ultimately enabling a quasi-direct observation of dynamic heterogeneities in liquid-cooled glasses.

## Acknowledgements

(JRV, CRT and MGS) and JF, JAP and JA acknowledge Grants PID2020-117409RB-I00, PGC2018-095032-B-100, PID2020-115663GB-C31 and PID2019-106744GB-I00, respectively, funded by MCIN/AEI/ 10.13039/501100011033. CRT is a Serra Hunter Fellow. The ICN2 is funded by the CERCA programme/Generalitat de Catalunya. The ICN2 was supported by the Severo Ochoa Centres of Excellence programme, funded by the Spanish Research Agency (AEI, grant no. SEV-2017-0706)".

## Author Contributions Statement.

JRV, CRT and MGS conceived the project. JF, MRR and AVC performed the AFM experiments. TB performed the OM measurements. TB, JF and MRR analysed the AFM and OM images. JAP and JA contributed with the mechanical analysis and interpretation. CRT, MGS and JRV discussed fundamental ideas of the transformation mechanisms. JRV wrote the manuscript with CRT, JF, JA, JAP and MGS. All authors commented on the manuscript. JRV supervised the project.

## Competing Interests Statement.

The authors declare no competing interests

## Methods

**Growth of organic glasses:** Previous experiments have shown that capping a low $T_g$ stable glass at both interfaces with a higher $T_g$ material inhibits the formation of liquid fronts at its surfaces/interfaces forcing the glass to devitrify on heating through a bulk-like mechanism[9,26,27,34,35]. The stack we use in this work consists of trilayers of TCTA/TPD/TCTA grown on a rigid substrate (native $SiO_2$/Si) or bilayers of TCTA/TPD/rigid substrate with various TPD and TCTA thicknesses. TCTA has a glass transition temperature of 428 K ($T_{g2}$) and the $T_g$ of TPD is 333 K ($T_{g1}$), so annealing above 333 K induces the transformation of the TPD while the TCTA remains as a rigid layer

Thin film glasses of TPD and TCTA were grown on both Si substrates and $SiN_x$ membranes at a substrate temperature of 0.85 $T_{g1}$ for TPD and 0.67 $T_{g2}$ for TCTA ($T_{sub}$=285 K), to ensure TPD is formed as ultrastable glass. Since the TPD thickness is only around 65 nm, it is capped at both sides by an ultrathin layer of a TCTA layer of 13 nm. The growth rate is fixed at 0.08 nm/s for both materials. The trilayer structure is schematically shown in figure 1a(left). The kinetic and thermodynamic stability of the TDP layers is measured before and after capping with ultrathin layers of TCTA by in-situ fast-scanning nanocalorimetry, as already shown elsewhere[35–37].

**In operando AFM:** Temperature-dependent AFM experiments were performed with a Keysight Technologies 5500 AFM controlled with a PicoScan 3000 electronics and enclosed in a chamber that was vibrationally and acoustically isolated[38]. Relative humidity (RH) was reduced below 10% RH by circulating nitrogen. Heating was performed by gluing the samples with silver paste to a modified commercial high temperature stage from Keysight Technologies. The temperature was controlled with a LakeShore 311 Temperature Controller in closed loop using the internal PT-100 sensor of the heater and measured with an additional PT-100 sensor glued with silver paste to the surface of the copper block of the heater (and close to the sample) and the resistance measured in the 4-wire configuration. The temperature at the sample surface was periodically calibrated with an additional PT-100 sensor glued with silver paste on a silicon wafer of the same thickness as that of the samples. The X, Y and Z dimensions of the piezoscanner were calibrated with reference samples. For the 1 um x 1 um scans, PS-PMMA block copolymers with fingerprint-like morphology and periods of 35 nm were used[39] while for larger scans (up to 20 um x 20 um) commercial calibration grating (TGXYZ01 and TGXYZ02 from MikroMasch) were used. Image were analysed with both WSxM[40] and Gwyddion[41].

The AFM was operated in open-loop intermittent contact mode using microfabricated silicon cantilevers with force constants of about 42 N m$^{-1}$ and ultrasharp silicon tips (nominal tip radius R < 10 nm) (PPP-NCHR, Nanosensors). The resonance frequencies were in the 250-350 kHz range. After laser alignment the sample was fixed to the heater and the heater stage coupled to the AFM head. When the whole setup was vibrationally and acoustically isolated several images ranging from 1 um x 1 um up to 20 um x 20 um scan size were taken until stabilization of the microscope. Once stabilized at room temperature, the sample was withdrawn, and the heater was driven up to $T_{g1}$-10 K at a heating rate of 5 K min$^{-1}$ (as recommended by the manufacturer). Once the desired temperature was reached, the resonance frequency was tuned (the frequency decreases for increasing temperature), the laser was aligned again, the tip was engaged, and few images were acquired to check for the stability of the instrument. Then, the sample was withdrawn again and the temperature increased up to the target temperature, $T_{g1}$+16 K. Again, when the temperature was reached, the resonance frequency was tuned not being necessary in most of the cases to adjust the laser position. Then the tip was engaged to find suitable regions to be able to follow the isothermal evolution of the surface as a function of time. When the temperature experiments were finished, the sample was withdrawn, the temperature control set to room temperature and after stabilization the resonance frequency was tuned, and the laser position corrected and few images at RT were acquired.

With the selected 512 points per line and 2 lines per second, each image took 256 s to complete (> 4 min). Lower acquisition times were tested resulting in poor resolution images. Although shorter times would be highly desirable, the evolution of the features was slow enough to be tracked satisfactorily. During withdrawal and engage cycles, the tip would not land exactly on the same position (due to hysteresis of the piezoscanner) which makes it difficult to localize a specific region in transformed samples.

At the target temperature, the apparent lateral dimensions (e.g., periodicity of the wrinkles) become smaller due to the increase of the strain coefficients (nm/V) of the piezoscanner with temperature which is not corrected during image acquisition[42]. The actual dimensions should be increased by a factor less than 10% (typically 5%), as obtained from images acquired at room temperature after the isothermal treatment. A similar trend is expected for the heights but in this case the correction is more involved since the tip/surface interaction should change with temperature, a behaviour that has not been explored sufficiently here, combined with the difficulty to localize a specific region at room temperature that was previously measured at high temperature, so that the obtained heights remain as apparent (uncorrected), although a 10% increase would be reasonable[42]. Further calibration experiments should be undertaken in order to quantify the variation of height with temperature.

**Finite element simulations:** The thin film system was modelled as an axisymmetric disc stiffened by the semi-infinite Si substrate (supplementary figure 3). The overall thin film geometry was discretized using 20-noded, brick-shape elements. Since possible processing stresses are unknown, the disc was assumed to be in the fully relaxed (or stress-free) configuration at the deposition temperature of 285 K. Linear thermoelasticity was then invoked as the temperature was ramped to 349 K in the first load step of the FE simulations. This temperature was set in accordance with that where the surface corrugation occurred in the experiments. An inner disc of radius $r$ representing the SCL region was then weakened in a second simulation stage, where numerical convergence was ensured after a slight temperature increment of 0.05 K. The finite element (FE) simulations were performed using ANSYS® Multiphysics (Release 19.0, http://www.ansys.com).

The cylindrical, weakened SCL regions were assumed to deform in accordance with linear elasticity with specific Young's modulus $E$ and Poisson's ratio $\nu$. Alternatively, the Neo-Hookean hyperelastic formalism was also invoked so as to account for the potential decrease of the elastic modulus of the weakened SCL regains at small strains. In this model, the total strain energy density is given by $U = C(I_1 - 3) + d^{-1}(\det \boldsymbol{F} - 1)^2$ where $C$ and $d$ are material parameters, $I_1$ is the first deviatoric strain invariant, and det$\boldsymbol{F}$ is the volume ratio. The second term in the right-hand side of the Neo-Hookean formulation explicitly accounts for possible material compressibility or volume changes, so that the SCL regions approach full incompressibility when $d \rightarrow 0$.

The capping TCTA top and bottom layers were assumed to remain linear elastic in all simulations, with matching Young's modulus $E$ and Poisson ratio $\nu$ to those in the glassy state (see the thermoelastic properties for the thin film materials and Si substrate in Supplementary Material). The material parameters for the SCL disc where then allowed to vary within the wider ranges of $C = 3.84 \times 10^6$ to $1.43 \times 10^7\ Pa$ and $d = 1.12 \times 10^{-7}$ to $2.79 \times 10^{-11}$. The experimentally observed transition between the onset of a single ridge to wrinkles with increasing $r$ from 250 to 1000 nm was then mimicked within the simulation subset where the SCL properties varied as previously.

**Data Availability**

Data that support the plots within this paper and other finding of this study are available from the corresponding authors upon reasonable request.

**Real-time microscopic view of the relaxation of an ultrastable glass**

M. Ruiz-Ruiz[1,2], A. Vila-Costa[1,2], Tapas Bar[2], Cristian Rodríguez-Tinoco[1,2], Marta Gonzalez-Silveira[1,2], Jose Antonio Plaza[3], Jorge Alcalá[4], Jordi Fraxedas[2], Javier Rodriguez-Viejo[1,2]

[1]*Departamento de Física. Facultad de Ciencias, Universitat Autònoma de Barcelona, 08193, Bellaterra, Spain*

[2]*Catalan Institute of Nanoscience and Nanotechnology (ICN2), CSIC and BIST, Campus UAB, Bellaterra, 08193, Barcelona, Spain*

[3]Instituto de Microelectrónica de Barcelona, IMB-CNM (CSIC), Esfera UAB, Campus UAB, Cerdanyola, Barcelona, Spain

[4]InSup, ETSEIB, Universitat Politècnica de Catalunya, 08028 Barcelona, Spain

**Amorphous nature of the layers after transformation**

We have verified that the isothermal treatments above $T_g$ do not induce the crystallization of the layers and therefore the observed surface corrugation are due to build-up stresses in the amorphous organic trilayer.

**(a)**                                    **(b)**

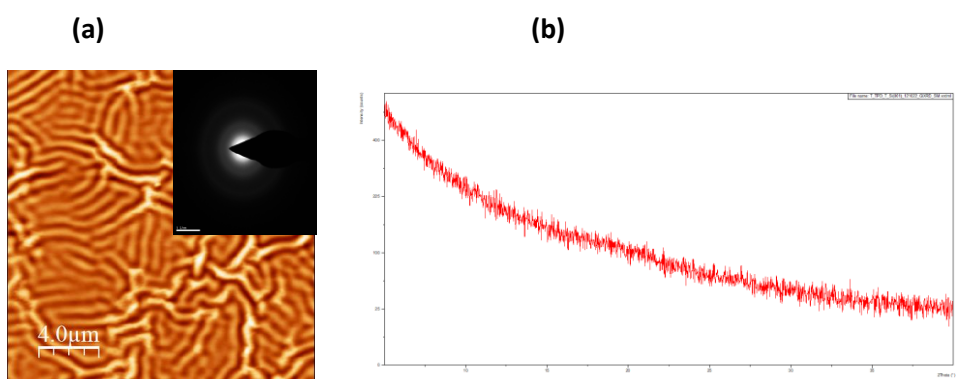

**Figure S1.** TCTA/TPD/TCTA sample (13/63/13 nm) (a) grown on a TEM grid and annealed in the AFM stage at Tg+21 K for 60 min. inset shows the electron diffraction pattern characteristic of a disordered sample. (b) X-ray diffraction profile of a sample grown on a Si substrate and annealed at Tg + 16 K to full transformation. The absence of any peak is also informative on the amorphous character of the trilayer.

**Wrinkling evolution without bottom TCTA layer**

This is a very informative experiment. We have previously verified that the native oxide/Si interface being weakly interacting with the TDP glass does not inhibit the appearance of liquid fronts at the interface. Therefore, a sample that is only capped by TCTA at the top will transform by planar fronts from the bottom interface when heated above $T_g$. This is relevant for the formation of wrinkles since the liquid forms at once at the whole surface and transformation is completed after few min at temperatures 16-20 K above Tg (TPD), due to the small thickness of the TPD layer, 63 nm. The wrinkled pattern appears from the fist moments at $T_{ann}$ which confirms our view that some shear transfer across the viscous liquid is necessary to induce surface corrugation.

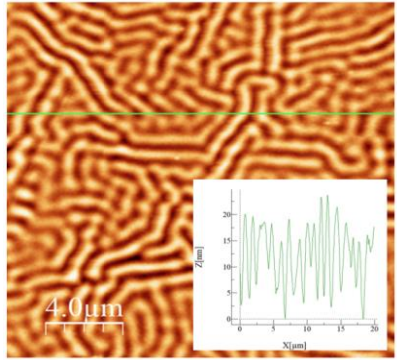

**Figure S2.** Corrugation pattern of a sample that transforms via front initiated at the interface with the substrate. AFM image taken at Tg + 21 K. Inset shows the cross-section profile.

**Mechanical analysis**

The geometry used for the finite element numerical simulations was cylindrical as schematically shown below.

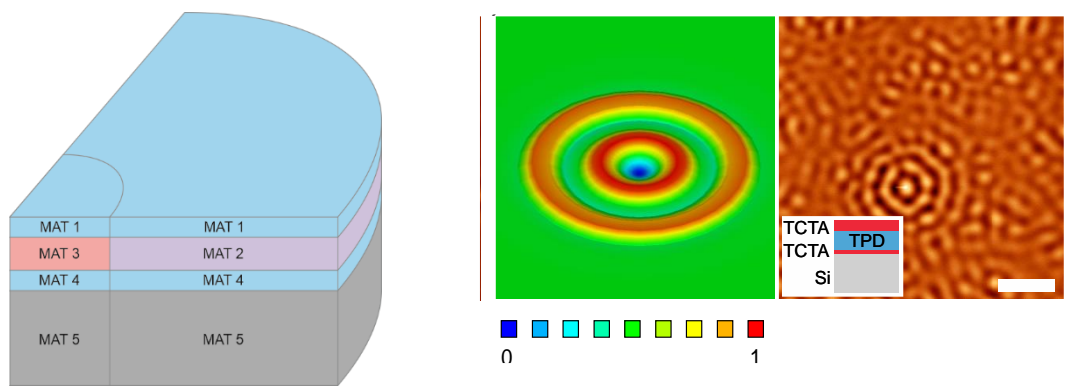

**Fig S3**. (a) Representative drawing of ¼ of the modelled structure showing the Materials distribution (MAT, top TCTA; MAT 2, solid TPD, MAT 3, fluidizes TPD; MAT 4. (b) concentric

wrinkles formed by changing the properties of the structure (FE) and doubling the thickness of the top TCTA layer, as 13/63/24 nm.

Table I. Material's properties

| Layer | Material nº | Thickness [nm] | Young's modulus E [Pa] | Poisson's ratio v | Coef. of Thermal Expansion |
|-------|-------------|----------------|------------------------|-------------------|----------------------------|
| TCTA (Top) | 1 | 13 | 5.4e9 | 0.3 | 2.3e-4 |
| TCTA (Buried) | 4 | 13 | 5.4e9 | 0.3 | 2.3e-4 |
| Silicon | 5 | >500 | 1.69e11 | 0.27 | 2.6e-6 |

| Layer | Material nº | Thickness [nm] | Initial Shear Modulus [Pa] | Incompressibility [Pa$^{-1}$] | Coef. Of Thermal Expansion |
|-------|-------------|----------------|----------------------------|-------------------------------|----------------------------|
| TPD (Solid) | 2 | 63 | 1.65e9 | 5.58e-10 | 2.3e-4 |
| TPD (Liquified) | 3 | 63 | 3.84E+06 to 1.43E+07* | 1.12E-07 to 2.79E-11* | 2.3e-4 |

* Equivalent ranges on Young´s Modulus and Poisson´s ratio for TPD solid and fluidified would be E = 4.3e+9 Pa and v = 0.3, and Y = [1.08E+07, 4.30E+07] and v = [0.4,0.4999], respectively. C = Initial shear modulus/2.

**Softening of the glass matrix during isotherms**

The ultrastable glass matrix slowly softens as a function of time. Evidence of this behaviour has been observed by monitoring through nanocalorimetry the glass transition of the ultrastable glass after different isotherms as shown in figure S4. The change in stability is minor compared to the ultrastability of the material. $T_{peak}$ changes by 5 K compared to $\Delta T \approx 40$ K with respect to the liquid-cooled glass. Interestingly, the partial softening does not play a significant role in the growth front velocity. At present its influence on the formation of liquid seed regions is still not completely clarified, (see main text).

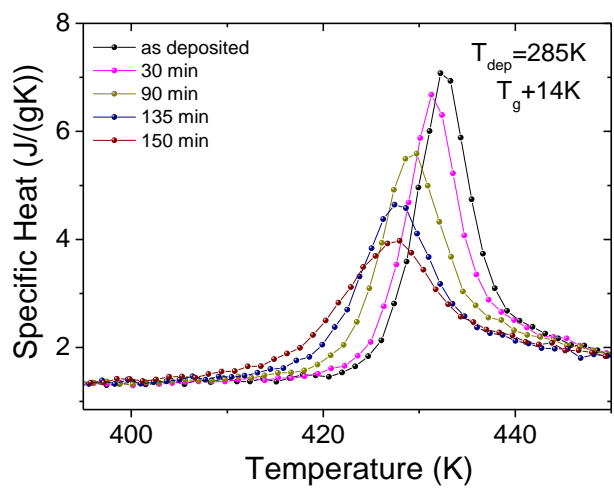

**Figure S4**. Calorimetric evidence of softening of the glassy matrix over time through the displacement of the endothermic overshoot associated to the glass transition of the ultrastable matrix during the annealing treatment. Data correspond to measurements at $T_g + 14$ K.

**Acceleration of the formation of new nuclei**

We give further details of the evaluation of the inset of figure 3a. We count the number of new nuclei that appear in each consecutive frame. Black circles represent this data normalized by the total area of each AFM image. The decrease in the number of nuclei as time evolves is a consequence of the reduction of the available surface due to transformation by dynamic facilitation into the liquid. Therefore, to account for this fact, we weight those values by the untransformed fraction, evaluated in figure 4b. The blue points represent the new nuclei per untransformed area. The 'nucleation' per untransformed area is very sensitive to the new nucleation that appears in each frame. However, there is a little mismatch in the count for the up and down images as the AFM trip scanned slightly less area during the up-to-down scanning. In the inset of figure 3a, we have considered an average of new nucleation numbers at the average time of two consecutive frames to get a true acceleration of nuclei that appear as time progresses.

**Evolution of growth velocity**

Growth front velocity evaluated from the slope of the radius versus time (figure S5) behaves linearly independent of the partial softening of the matrix in a range of isothermal temperatures.

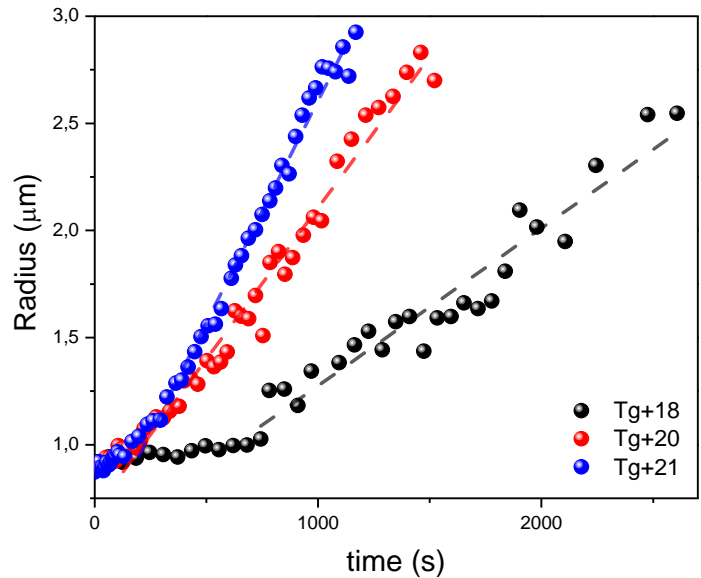

**Figure S5**.
Measurements on ultrastable TDP glass grown at 0.85 $T_g$. (a) Growth of liquid regions as a function of time obtained from optical microscopy images at three annealing temperatures.

**Growth front velocity of an ultrastable TPD glass: nanocalorimetry measurements**

The growth front velocity at high T using fast scanning nanocalorimetry is evaluated as explained previously [1]. Non-capped ultrastable TPD layers are used. The low temperature point at Tg + 14 K is obtained by analyzing the transformed fraction after different isotherms. The data obtained in the present work through AFM and optical microscopy (red dots) fit well to the TPD data (dashed line).

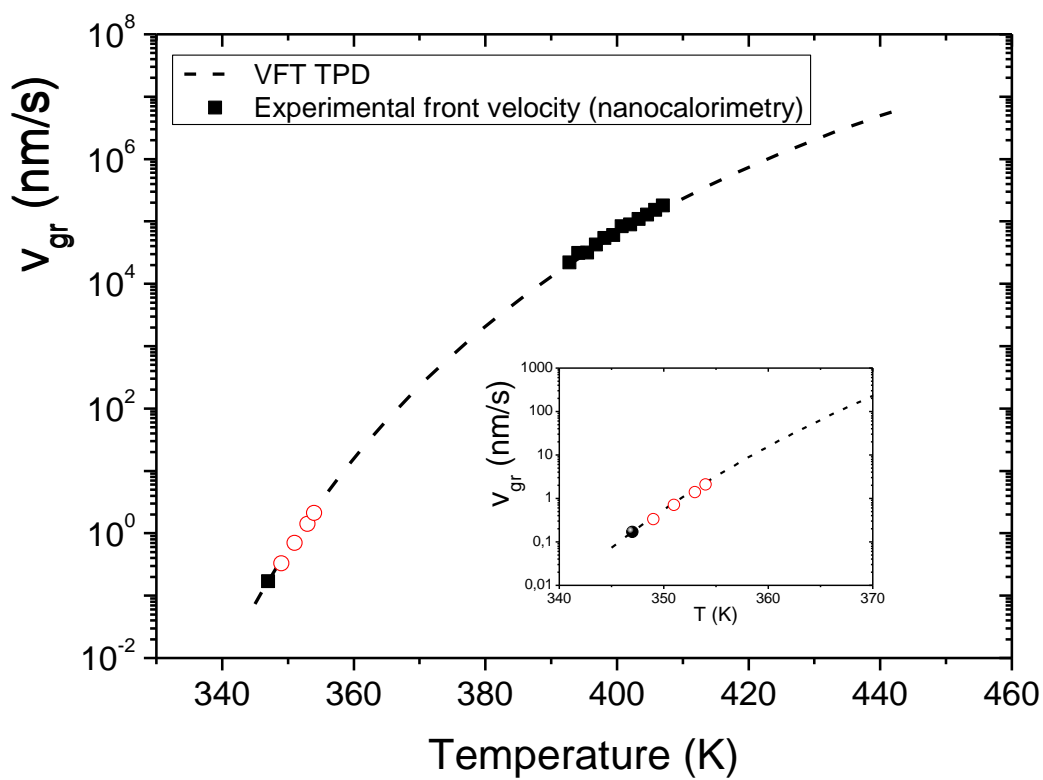

**Figure S6.** Growth front velocity obtained from independent nanocalorimetric measurements on non-capped TPD films. High T points by fast scanning of the as-deposited and low-$t$ by isotherms at Tg + 14 K. The dashed line is the $v = C \cdot \tau^{-\gamma}$ for TPD, being τ the alpha relaxation time. (inset): amplified temperature region of interest. Red circles correspond to experimental data of the growth velocity of the liquid regions obtained from AFM (OM) images at 4 different temperatures.

**Influence of annealing temperature and thickness**

Although the main text focuses on 13/63/13 nm samples annealed at $T_g$ + 16K we have measured a large number of samples of different thicknesses (60-to-400 nm) and at different annealing temperatures spanning Tg+16 to Tg+21 K. While there are slight differences between them the mechanism of transformation remains the same.

Figure S7 shows a series of AFM images of a capped trilayer (TCTA/TPD/TCTA) but with higher TDP thickness (13 nm/200 nm/13 nm) during the transformation at $T_g$ + 20 K and figure S8(a) show details of one specific wrinkled pattern and the line profiles as it grows over time and Fig. S8(b) a cross-section SEM image obtained after fracturing a totally transformed sample to show the trilayer structure and the presence of surface undulations without any detachment from the substrate. The rough appearance is due to the 5 nm of Pt that was used for metallization.

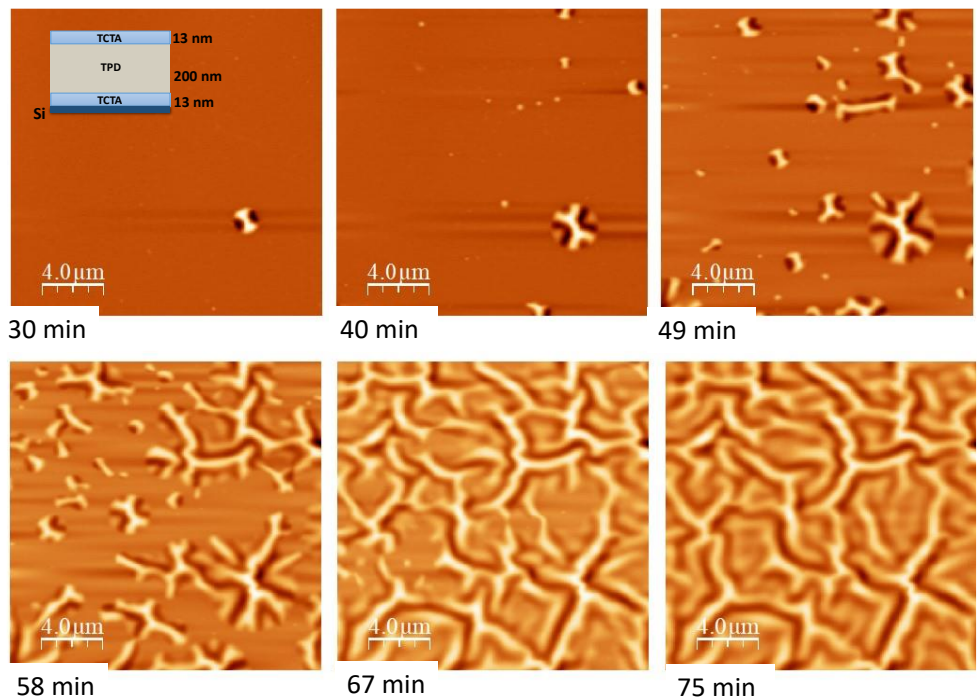

**Figure S7**. AFM scans obtained during an isotherm at $T_g$ + 20 K after different times. Full transformation is achieved after 75 min.

**(a)**

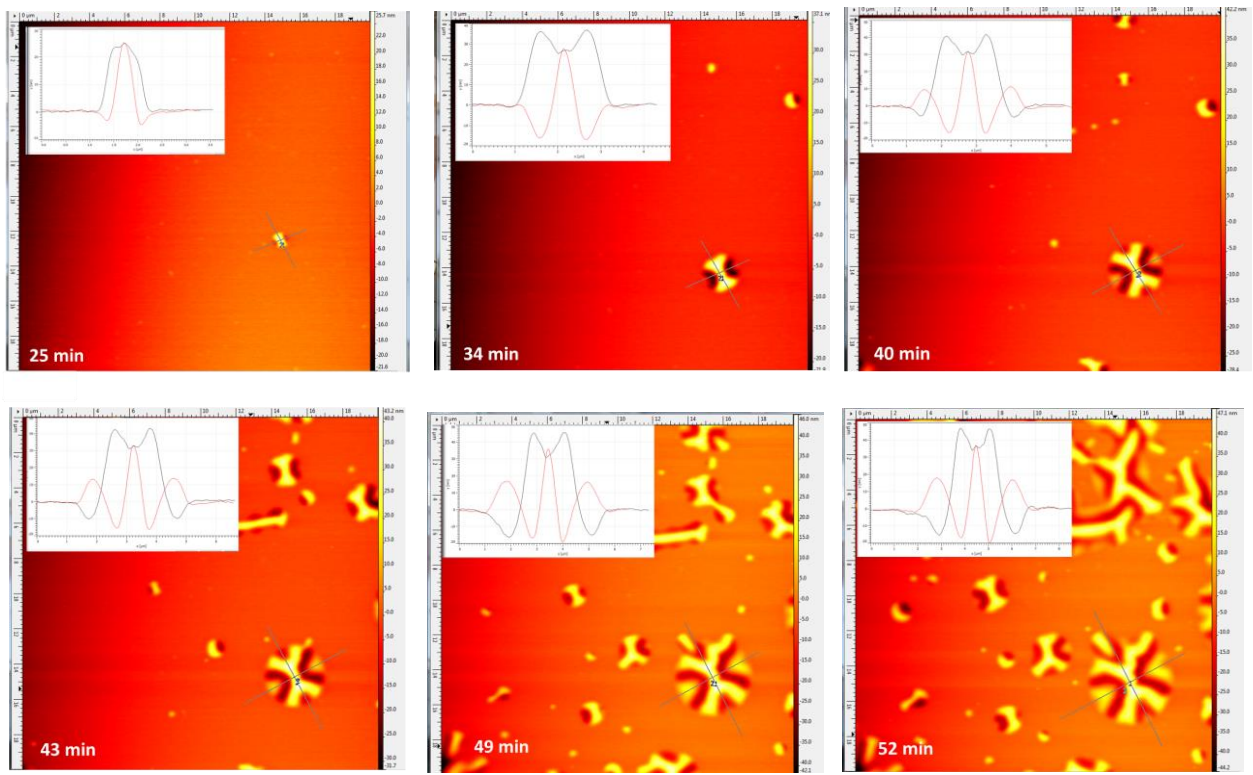

**(b)**

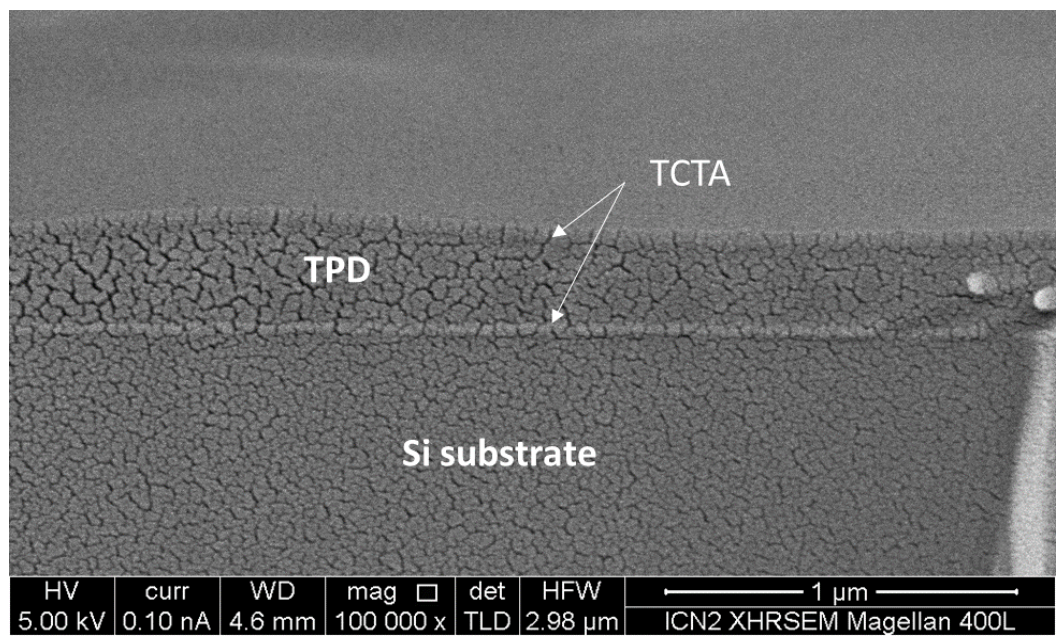

**Figure S8**. (a) AFM scans and line profiles of a circular pattern that grows radially over time. (b) Cross section scanning electron micrograph of the undulations in a 13/200/13 nm sample.

**Impact of density of 'surface defects' on the transformation**

The roughness of the surface has a substantial effect on the nucleation behavior nature of liquid regions in the middle TPD interlayer. The transformation time of a sample presenting dispersed small hillocks (2-4 nm high) on the $SiO_2$/Si surface is reduced with respect to one with a clean surface. The dynamics is, at least partially, driven by these imperfections. Some of them are preferential sites for liquid formation accelerating the dynamics while others remain inactive for long periods of time. The reason is at present unknown. On the contrary, the lateral growth velocity of the liquid is independent of the initiation site of equilibration. This sensitivity to surface defects (and substrate roughness) (Fig. S10) may explain the different transformation dynamics reported previously in the same system (*34*), where the transformed fraction obtained by nanocalorimetry was fitted with a small Avrami exponent of two, suggesting the presence of preexisting nucleation sites that drove the formation of liquid regions.

Figure S10 shows two samples of ultrastable TPD sandwiched by TCTA 20/200/20 (TCTA/TPD/TCTA). Blue symbols correspond to a trilayer grown on a clean $SiO_2$/Si surface (roughness RMS in 1x1 $\mu m^2 \approx 0.13$ nm) while red symbols correspond to an identical trilayer grown on a rougher substrate. Measurements are carried out by OM (see images) at $T_g + 16$ K. The rougher substrate induces a faster formation of liquid regions.

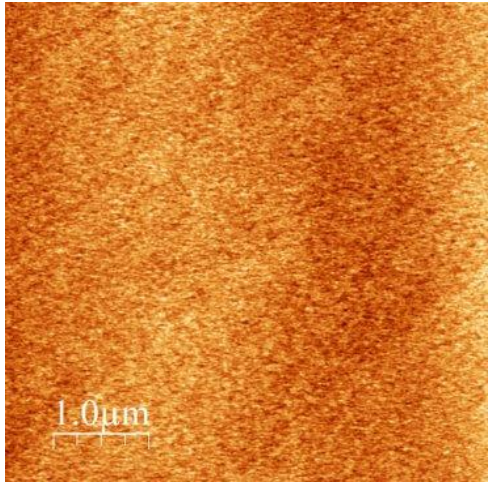
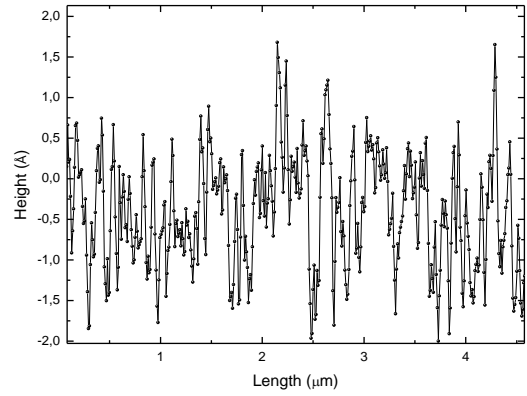

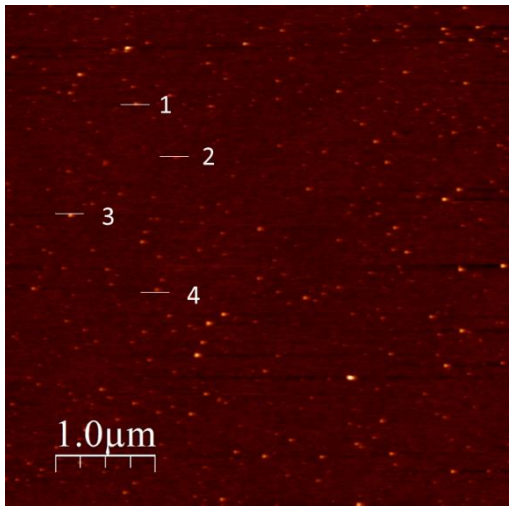
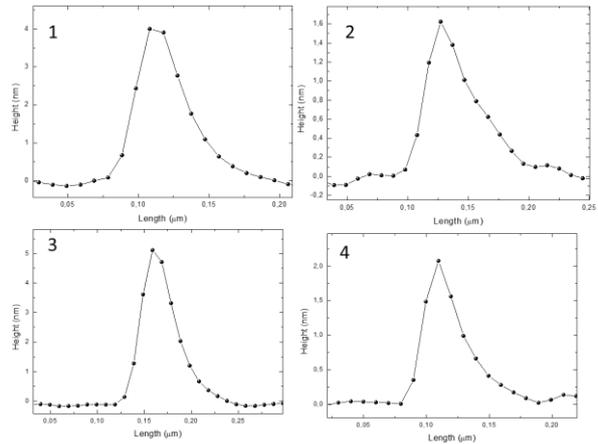

**Figure S9.** (top) AFM image and roughness profile of a 10x10 µm² clean native oxide/Si surface. (bottom) AFM image of a native oxide/Si surface with many surface defects that induce accelerated dynamics during the transformation of the TPD in the trilayer structure and line profiles of some defects marked in the AFM image.

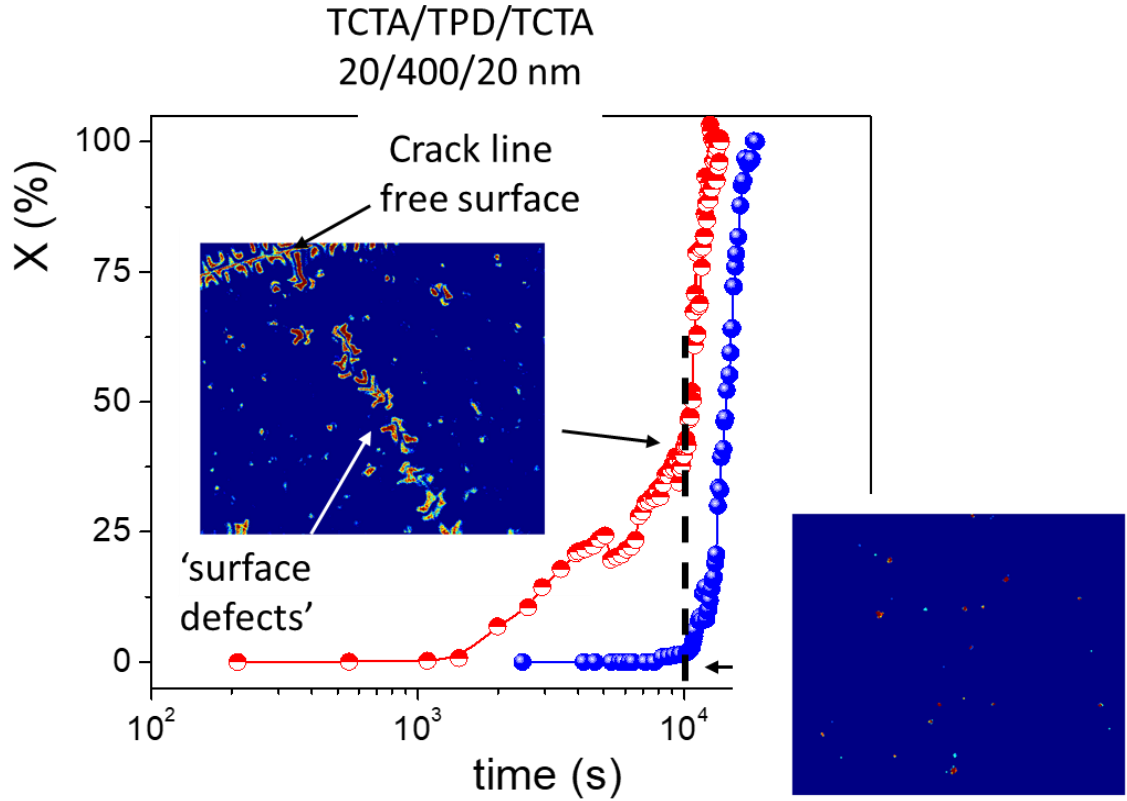

**Fig. 10.** The transformation time of a sample presenting dispersed small hillocks (3-4 nm height on average) on the $SiO_2$/Si surface (red circles) is reduced with respect to one grown on a clean surface (blue circles). As observed in the figure the formation of liquid seeds is enhanced in the sample grown on rougher substrate. Representative optical microscopy images showing preferential formation of liquid regions along a line of defects (centre) by front in a crack (superior left corner) and another sample with a clean surface after 10200 s at $T_{g1}$ + 16 K. The thickness of this trilayer was TCTA (20 nm)/TDP (400 nm)/TCTA (20 nm) and it partially cracked during the annealing. The images are treated by subtracting the background of a clean image and colouring the wrinkles to aid visualization.